\documentclass[aps,prl,twocolumn,showpacs,preprintnumbers,superscriptaddress]{revtex4}

\usepackage{amsmath}
\usepackage{amssymb}
\usepackage{graphicx}
\usepackage{dcolumn}
\usepackage{bm}
\newcommand{\YP}{\rm Yb_{3}Pt_{4}}

\begin{document}

\title{Magnetic field tuning of antiferromagnetic Yb$_{3}$Pt$_{4}$}

\author{L. S. Wu}
\affiliation{Department of Physics and Astronomy,
Stony Brook University, Stony Brook, New York 11794-3800, USA}
\author{Y. Janssen}
\affiliation{Condensed Matter Physics and Materials Science
Department, Brookhaven National Laboratory, Upton, New York
11973-5000, USA}
\author{C. Marques}
\affiliation{Department of Physics and Astronomy,
Stony Brook University, Stony Brook, New York 11794-3800, USA}
\author{M. C.  Bennett}
\affiliation{Condensed Matter Physics and Materials Science
Department, Brookhaven National Laboratory, Upton, New York
11973-5000, USA}
\author{M. S. Kim}
\affiliation{Department of Physics and Astronomy,
Stony Brook University, Stony Brook, New York 11794-3800, USA}\affiliation{Condensed Matter Physics and Materials Science
Department, Brookhaven National Laboratory, Upton, New York
11973-5000, USA}
\author{K.  Park}
\affiliation{Condensed Matter Physics and Materials Science
Department, Brookhaven National Laboratory, Upton, New York
11973-5000, USA}
\author{Songxue Chi}
\affiliation{NIST Center for Neutron Research, NIST, Gaithersburg, Maryland 20899, USA}\affiliation{Department of Materials Science and Engineering, University of Maryland, College Park, Maryland 20742, USA}
\author{J. W. Lynn}
\affiliation{NIST Center for Neutron Research, NIST, Gaithersburg, Maryland 20899, USA}
\author{G. Lorusso}
\affiliation{CNR-Institute of nanoSciences S3, via G. Campi 213/a, 41100 Modena, Italy}\affiliation{Dipartimento di Fisica, Universit$\grave{a}$ di Modena e Reggio Emilia, via G. Campi 213/a, 41100
Modena, Italy}
\author{G. Biasiol}
\affiliation{Laboratorio Nazionale TASC-INFM, Area Science Park Basovizza, I-34012 Trieste, Italy}
\author{M. C. Aronson}
\affiliation{Department of Physics and Astronomy,
Stony Brook University, Stony Brook, New York 11794-3800, USA}\affiliation{Condensed Matter Physics and Materials Science
Department, Brookhaven National Laboratory, Upton, New York
11973-5000, USA}
\date{\today}

\begin{abstract}
We present measurements of the specific heat, magnetization, magnetocaloric effect and magnetic neutron diffraction carried out on single crystals of antiferromagnetic Yb$_{3}$Pt$_{4}$, where highly localized Yb moments order at $T_{\rm N}=2.4$ K in zero field. The antiferromagnetic order was suppressed to $T_{\rm N}\rightarrow 0$ by applying a field of 1.85 T in the $ab$ plane. Magnetocaloric effect measurements show that the antiferromagnetic phase transition is always continuous for $T_{\rm N}>0$, although a pronounced step in the magnetization is observed at the critical field in both neutron diffraction and magnetization measurements. These steps sharpen with decreasing temperature, but the related divergences in the magnetic susceptibility are cut off at the lowest temperatures, where the phase line itself becomes vertical in the field-temperature plane. As $T_{\rm N}\rightarrow0$, the antiferromagnetic transition is increasingly influenced by a quantum critical endpoint, where $T_{\rm N}$ ultimately vanishes in a first order phase transition.
\end{abstract}


\maketitle

\section{Introduction}

Materials where magnetic order can be suppressed to low or even vanishing temperatures have proven to be rich sources of new physics. In different families of compounds, based both on transition metal and rare earth moments, the relative weakness or absence of competing magnetic phases makes it possible to observe new types of ordered states, most notably superconductivity~\cite{mathur1998,saxena2000} and quasi-ordered phases such as `spin nematics'~\cite{borzi2007}, that would normally be obscured. The magnetic excitations are greatly modified when the onset of magnetic order occurs at low temperatures, due to the importance of quantum mechanical fluctuations between the ordered and disordered states, leading to their characteristic E/T scaling ~\cite{aronson1995,schroder2001} and to unusual temperature divergencies in the specific heat and magnetic susceptibility~\cite{coleman2005,stewart2001,vonlohneysen2007,vonlohneysen2008,gegenwart2008,steglich2010}. It is a matter of continuing debate as to how these fluctuations enable or destabilize novel orders, for instance whether they provide a pairing mechanism for unconventional superconductors~\cite{monthoux2001}.

Very few compounds form with magnetic order restricted to zero temperature, and in most cases it is necessary to use pressures, compositions, or magnetic fields to tune the ordering temperature to $T=0$ to form a quantum critical point (QCP) if magnetic order is continuous, or a quantum critical end point (QCEP) if the magnetic transition becomes first order. It is well appreciated that quantum critical compounds are exquisitely sensitive to disorder, and it has been established that even modest amounts of disorder can change the order of magnetic transitions if the transition temperature is sufficiently low~\cite{belitz2005,sokolov2006,uhlarz2004}. Pressure tuning of magnetic transitions has an appealing simplicity, since it largely avoids these concerns about disorder,  but experimental access is somewhat limited, due to the bulky equipment needed for high pressure measurements. Thermodynamic measurements are especially problematic at high pressures, although they are of particular value for understanding how cooperative phases are stabilized at the lowest temperatures. For these reasons, magnetic field tuning of magnetic transitions is increasingly attractive, although it has been noted that the quantum criticality induced by field and pressure within a single material may not be identical~\cite{lohneysen2001,fischer2005,berry2010}. Magnetic fields affect the stability of magnetic order  at two different levels. First, fields can destabilize the magnetic structure, selected by the system as the lowest energy configuration for $T\rightarrow 0$ in zero field. This is effected by the suppression of critical fluctuations, hampering the establishment of long-ranged and long lived magnetic correlations that can lead to magnetic order itself.  Second, magnetic fields can change the properties of individual magnetic moments as well, resulting in Zeeman splitting of the states of the crystalline electric field manifold, and in some cases by the suppression of moment compensation by the Kondo effect. Both effects are expected to be important for heavy fermion compounds, where two limiting behaviors can be identified. In one case, magnetic order emerges at $T_{\rm N}$  from a paramagnetic state where the moments are highly localized, having only a weak exchange coupling to the conduction electron states whose energy scale $k_{\rm B}T_{\rm 0}\leq k_{\rm B}T_{\rm N}$. Alternatively, the crystal field states can be extensively broadened via hybridization, possibly to the point that the localized character can be considered minimal or absent when magnetic order occurs at $k_{\rm B}T_{\rm N}\leq k_{\rm B}T_{\rm 0}$.

Field tuning experiments have been extensively pursued in complex systems like CeCu$_{6-x}$Au$_{x}$~\cite{lohneysen2001} and  YbRh$_{2}$Si$_{2}$~\cite{custers2003}, where the antiferromagnetic phase line remains continuous as $T_{\rm N}\rightarrow 0$ at a quantum critical field $B_{\rm QCP}$. It is evident here that not only does the magnetic order evolve with field, but also the underlying electronic structure can itself be critical at or near $B_{\rm QCP}$~\cite{gegenwart2008,friedemann2009}.  We present here an experimental study of the field-temperature phase diagram of the heavy fermion antiferromagnet Yb$_{3}$Pt$_{4}$. $\YP$ orders antiferromagnetically at a N\'{e}el temperature $T_{\rm N}=2.4$ K~\cite{bennett2009}. While $\YP$ is metallic, magnetic order develops directly from a paramagnetic state where the fluctuating moments correspond to the ground doublet of the crystal field split Yb$^{3+}$ ion, with no indication of any Kondo effect. The spin waves in the antiferromagnetic state are conventional, resulting from the exchange field acting on the doublet ground state~\cite{aronson2010}. The unit cell of $\YP$ is very large, and the absence of strong magnetic anisotropy suggests that Fermi surface nesting will play little role here in stabilizing magnetic order.  We will argue here that the relative simplicity of the antiferromagnetic order in $\YP$ allows us to explore the field tuning of antiferromagnetic order without the complexities of electronic delocalization that are found in systems like YbRh$_{2}$Si$_{2}$.

We present the results of specific heat, magnetization, and magnetic neutron diffraction measurements that demonstrate that magnetic fields suppress antiferromagnetic order in $\YP$, causing it to vanish at a critical field of 1.85 T.  An analysis of the magnetization, specific heat, and the magnetocaloric effect indicates that the magnetic order remains continuous for all nonzero temperatures, but that the influence of a $T=0$ QCEP becomes increasingly strong as $T_{\rm N}\rightarrow 0$, leading to qualitative modifications to the phase line  when $T_{\rm N}\leq 1.2$ K. Divergences in the temperature dependence of the magnetic susceptibility are cut off at the lowest temperatures, suggesting that antiferromagnetic order in $\YP$ occurs via a first order transition at zero temperature. The field-temperature phase diagram found for $\YP$ is of a type that has not been previously reported for heavy fermion compounds, although it combines features of metamagnets and also systems with true first order transitions.

\section{Experimental Details}

Single crystals of Yb$_{3}$Pt$_{4}$ were grown from lead flux, and powder x-ray diffraction measurements were used to verify the rhombohedral Pu$_{3}$Pd$_{4}$ structure type~\cite{bennett2009,janssen2010}. The field $B$ and temperature $T$ dependent dc magnetization $M(B,T)$ was measured using a Quantum Design Magnetic Properties Measurement System (MPMS) for temperatures above 1.8 K, and at lower temperatures using a Hall sensor-based technique that was calibrated to the MPMS data above 1.8 K~\cite{neil, Cavallini2004, Candini2006}. The specific heat was measured for temperatures that ranged from 0.1 K to 4 K, and in fields as large as 3 T using a Quantum Design Physical Property Measurement System (PPMS), equipped with $^{3}$He and dilution refrigerator inserts. As described elsewhere, the magnetic and electronic parts of the specific heat $C_{\rm M}$ were obtained by subtracting the specific heat of isostructural but nonmagnetic Lu$_{3}$Pt$_{4}$ from the total specific heat $C$~\cite{janssen2010}. Measurements of the magnetocaloric effect (MCE) were performed using the PPMS specific heat puck, where the sample was heat sunk to a calibrated resistive temperature sensor. These experiments were carried out in the adiabatic limit, as the field sweeps were significantly faster than the measured thermal relaxation time of the pucks~\cite{rost2009}. Neutron diffraction was carried out on a 65 mg single crystal of $\YP$ at the NIST Center for Neutron Research using the BT- 7 double focusing triple-axis spectrometer with the neutron wavelength $\lambda=2.47$ $\AA$.

\section{Experimental Results}

X-ray diffraction shows that Yb$_{3}$Pt$_{4}$ crystallizes in the reported rhombohedral Pu$_{3}$Pd$_{4}$-type of structure~\cite{palenzona1977}, which has 18 Yb atoms per unit cell, all with the same site symmetry, and 24 Pt atoms per unit cell, with three different site symmetries. $\YP$ orders antiferromagnetically at the N\'{e}el temperature $T_{\rm N}=2.4$ K~\cite{bennett2009} and the magnetic structure was determined from neutron diffraction measurements using representation analysis~\cite{janssen2010}. The fundamental building block of this $q=0$ antiferromagnetic structure is a triad of Yb moments, each rotated 120 degrees with respect to each other. Each triad is matched by a reflected triad to form octahedra, which are stacked in a staggered fashion along the $\emph{c}$-axis to form the overall magnetic structure. Magnetization measurements indicate that the hard axis is along the $\emph{c}$-axis, and the easy axis lies in the $ab$ plane. The magnetic anisotropy is weak inside the $ab$ plane, with $\chi_{\rm [110]}/\chi_{\rm [100]}\simeq 1.07$. It is much bigger between the $ab$ plane and $c$ axis, with $\chi_{\rm ab}/\chi_{\rm c}\simeq 6$ at low temperatures.

There is significant evidence that the Yb moments in $\YP$ are spatially localized over much of the range of experimental temperatures, and so their $f$ electrons are excluded from the metallic Fermi surface. The magnetic susceptibilities for fields along both the $c$ axis and in the $ab$ plane are in agreement with Curie-Weiss expressions above $\simeq$ 150 K, giving a paramagnetic moment about 4.24 $\mu_{B}$/Yb, as expected for trivalent Yb~\cite{bennett2009}. A pronounced anomaly in the zero field specific heat $C$ is well described by a Schottky expression involving four crystal-field-split doublets, just as expected for Yb$^{3+}$ in a crystal symmetry that is lower than cubic. Inelastic neutron scattering measurements confirm that there are four magnetic doublets that are well separated in energy, and since the first excited state is $\simeq7.5$ meV ($\sim87$ K) above the ground state~\cite{janssen2010}, this ground doublet dominates the magnetic properties of $\YP$ at low temperatures.  Antiferromagnetic order occurs in $\YP$ at 2.4 K, signalled by a mean field peak in the specific heat~\cite{bennett2009}. The entropy reaches $\sim 0.8$ Rln2 at $T_{\rm N}$, confirming that the doublet moment orders with a  minimum of critical fluctuations or with appreciable suppression of the ordering moment via the Kondo effect. Triple axis spectroscopy was used to show that the temperature evolution of the spin waves in the antiferromagnetic state~\cite{aronson2010} is similar to that of the magnetic order parameter, suggesting that the spin waves are conventional and arise from the action of the exchange coupling on the crystal field split single ion states.

\begin{figure}
\includegraphics[width=6.50cm]{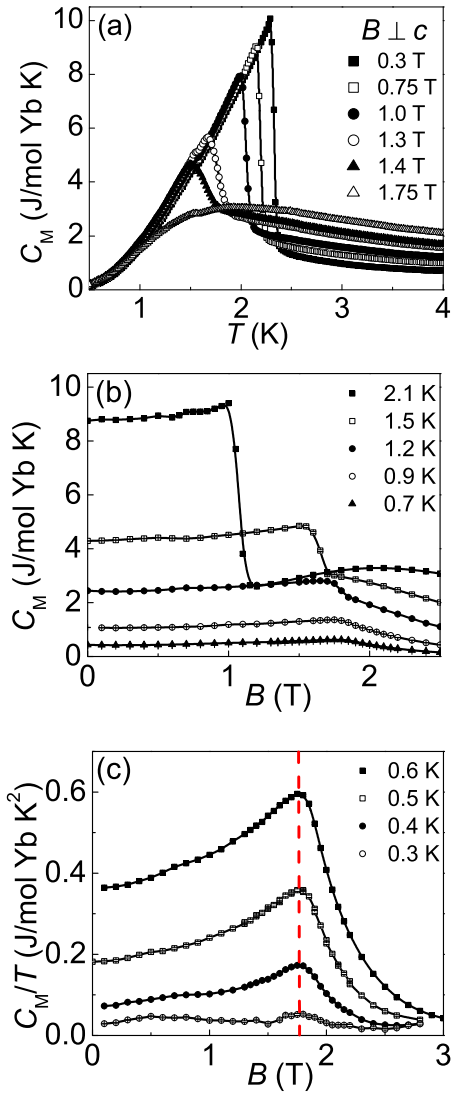}
\caption{(Color online) (a) Temperature dependencies of the magnetic and electronic specific heat $C_{\rm M}$. (b) Field dependencies of $C_{\rm M}$ at different fixed temperatures, as indicated. (c) Field dependencies of $C_{\rm M}/T$ at different fixed temperatures, as indicated. Dashed line shows that the phase line $T_{\rm N}(B)$ becomes field independent for $T_{\rm N}\leq 0.9$ K. The magnetic field in (a)-(c) is perpendicular to the $c$ axis. Solid lines in (a)-(c) are guides for the eye. \label{fig_CvsTH}}
\end{figure}

We have measured the temperature dependencies of the magnetic and electronic specific heat $C_{\rm M}$ of $\YP$ with different values of the magnetic field $B$ in the $ab$ plane Fig.~(\ref{fig_CvsTH}(a)). Since the magnetic anisotropy inside the $ab$ plane is very small, we do not specify the magnetic field direction inside the $ab$ plane for all the experiments showing here and below. In low fields, the specific heat jump at $T_{\rm N}$ has a triangular shape evocative of a mean-field transition. $T_{\rm N}$ decreases with increasing field, while the magnitude of the ordering anomaly decreases and eventually becomes undetectable for fields greater than $\simeq 1.75$ T, where $T_{\rm N}<1.2$ K. While these data may suggest that the antiferromagnetic phase line $T_{\rm N}(B)$ terminates at a critical endpoint with $T_{\rm N}=1.2$ K, $B=1.75$ T, it is also possible that it simply becomes very steep as $T_{\rm N}\rightarrow 0$. To distinguish between these two possibilities, field scans of the specific heat $C_{\rm M}(B)$ were performed at different fixed temperatures (Fig.~\ref{fig_CvsTH}(b)). Very different behaviors were found above and below 1.2 K. For $T \geq 1.2$ K,  there is a step in $C_{\rm M}(B)$ as the field transits the phase line $T_{\rm N}(B)$, reminiscent of the step that is found in $C_{\rm M}(T)$ when increasing temperature is used to suppress antiferromagnetic order in a fixed magnetic field (Fig.~\ref{fig_CvsTH}(a)). This step evolves into a broad peak centered at $T_{\rm N}(B)$ for $T \leq 1.2$ K, whose magnitude decreases and becomes very small at the lowest temperatures (Fig.~\ref{fig_CvsTH}(c)). There is no measurable change in the field at which the peak in $C_{\rm M}(B)$ occurs for any temperature below $\simeq 0.9$ K, indicating that within the accuracy of our  measurements the antiferromagnetic phase line becomes vertical in the $B-T$ plane as $T_{\rm N}\rightarrow 0$ for the magnetic field $B_{\rm 0}=1.85$ T.

\begin{figure}
\includegraphics[width=7.50cm]{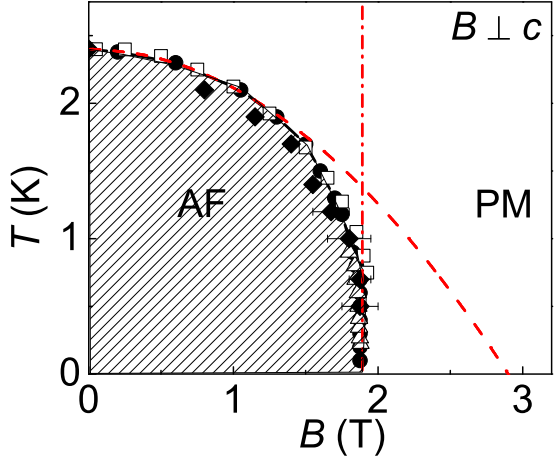}
\caption{(Color online) Antiferromagnetic order is found in the shaded area of the field -temperature phase diagram of $\YP$, where the phase boundary $T_{\rm N}(B)$ is determined from field scans of the specific heat $C_{\rm M}$ ($\bullet$), from the temperature dependencies of the magnetization M, carried out in different fixed fields ($\square$), from the field dependencies of the magnetization M, carried out at different fixed temperatures ($\triangle$), and from the magnetic intensity of the (110) Bragg peak, measured in a neutron diffraction experiment for different fixed temperatures and fields ($\blacklozenge$). Error bars indicate the width of the moment step in the neutron diffraction experiment. The red dashed line is a fit to the expression $T_{\rm N}(B)= T_{\rm N}(0)[1-(\frac{B}{B_{\rm 0}})^{2}]$, where $T_{\rm N}(0)=2.4$ K and $B_{0}=2.9$ T. Vertical dash - dot line indicates the 1.85 T field at which $T_{\rm N}\rightarrow 0$. Solid line is a guide for the eye. \label{fig_HT_phase_diagram}}
\end{figure}

The full antiferromagnetic phase line $T_{\rm N}(B)$ determined from field sweeps of the specific heat $C$ is presented in Fig.~\ref{fig_HT_phase_diagram}. At the lowest fields, $T_{\rm N}(B)$ follows a smooth power-law from its $B=0$ value $T_{\rm N}(0)=2.4$ K, i.e. $T_{\rm N}(B)= T_{\rm N}(0)[1-(\frac{B}{B_{\rm 0}})^{2}]$, qualitatively consistent with the mean-field nature of the phase transition found in this part of the phase diagram and suggesting a conventional quantum critical point at a field $B_{\rm 0} \simeq 2.9$ T that is never actually reached.  The phase line abruptly deviates from this behavior as the field approaches 1.85 T, and since its final approach to the $T=0$ axis cannot be described by any power law, quantum criticality is ultimately avoided in field tuned $\YP$.

\begin{figure}
\includegraphics[width=6.5cm]{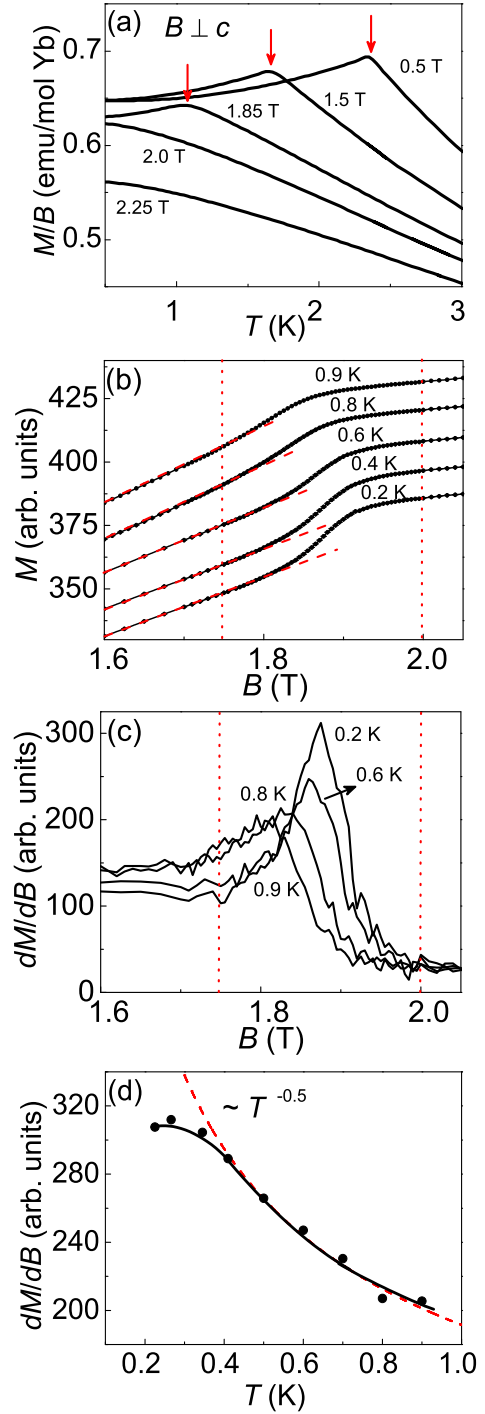}
\caption{(Color online) (a) Temperature dependencies of the magnetization $M$ divided by different measuring fields $B$, as indicated. The arrows indicate the antiferromagnetic transitions in each field. (b) Field dependencies of $M$ at indicated temperatures. Vertical dotted lines indicate the range of fields where $M(B)$ deviates from linearity, as evidenced by departures from linear extrapolations of low field $M(B)$ (red dashed lines).  (c) The numerical derivative $\chi=dM/dB$ of the data in (b). (d) Temperature dependence of the maximum value of $\chi=dM/dB$ from (c). The dashed line is the fit to $\chi\simeq T ^{\rm -0.5}$, while the solid line is guide for the eye that emphasizes the saturation of $\chi$ for $T\leq 0.35$ K. In (a)-(c), the magnetic field is perpendicular to the $c$ axis.\label{fig_Magnetization}}
\end{figure}

A more detailed picture of the antiferromagnetic phase transition is revealed by the magnetization measurements presented in Fig.~\ref{fig_Magnetization}. The temperature dependencies of the magnetization $M/B$ were measured in different fixed fields $B$ (Fig.~\ref{fig_Magnetization}(a)), displaying distinct cusps at $T_{\rm N}$. As we found in the specific heat measurements, $T_{\rm N}$ is driven to lower temperatures by the application of magnetic fields $B$, and the values of $T_{\rm N}(B)$ agree very well between the two measurements (Fig.~\ref{fig_HT_phase_diagram}).  The ordering anomaly in $M(T)/B$ broadens and is no longer observed above 0.5 K for $B\geq 1.85$ T. Given the vertical nature of the phase line $T_{\rm N}(B)$ revealed by the specific heat measurements, we turn to field sweeps of the magnetization $M(B)$ to clarify the phase behavior at the lowest temperatures. We emphasize that no hysteresis is observed between measurements performed with increasing and decreasing fields, at any field or temperature. Fig.~\ref{fig_Magnetization}(b) shows that $M(B)$ is initially linear in field, but deviates from this initial slope near the field-driven transition at 1.85 T before becoming linear again with a much smaller slope at the highest fields.  With decreasing temperature, this slope change becomes sharper, suggesting that the associated differential susceptibility $\chi(B)=dM/dB$ is becoming very large at $T_{\rm N}(B)$. Indeed, Fig.~\ref{fig_Magnetization}(c) shows that there is a distinct peak in $\chi(B)$ that becomes sharper and increases strongly in magnitude as the temperature decreases. Figure~\ref{fig_Magnetization}(d) shows that the maximum value of the susceptibility $\chi$ at $T_{\rm N}(B)$ initially increases according to a power law $\chi \sim T^{-1/2}$, but saturates below $\simeq 0.35$ K. We considered the possibility that experimental factors may play a role in this saturation, for instance the degree of thermal sinking of the sample on the Hall sensor, found to be appreciable below $\simeq 0.15$ K, as well the precision of the $M(B)$ measurement itself, which limits the degree of divergence possible in $\chi(B)$, obtained by numerically differentiating $M(B)$. These effects are minimal above 0.2 K, where the saturation of the power law divergence of $\chi(T)$ primarily reflects a broadening of the antiferromagnetic transition, due either to disorder in the sample or alternatively by thermal or quantum fluctuations. It is evident from Fig.~\ref{fig_Magnetization}(b) that the width of $\chi(B)$ decreases slightly with decreasing temperature, suggesting that disorder is not the only factor determining the breadth of the field driven transition as $T_{\rm N}\rightarrow 0$, but that quantum fluctuations are also likely to play an increasing role.

\begin{figure}
\includegraphics[width=6.50cm]{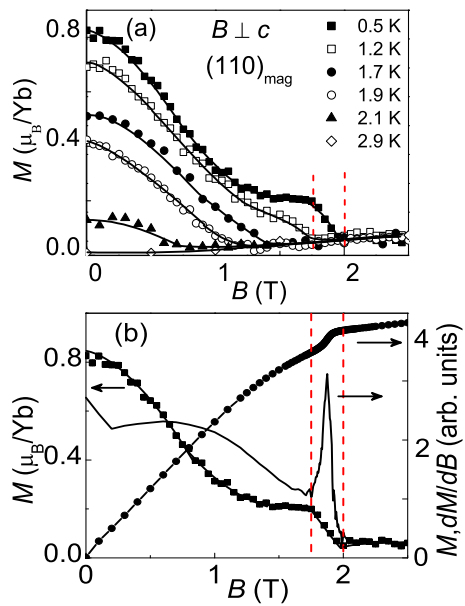}
\caption{(Color online) (a)Field dependencies of the (110) magnetic Bragg peak intensity at different temperatures, as indicated. Solid lines are guides for the eye. (b) The magnetization($\bullet$), and differential magnetic susceptibility (solid line) at 0.2 K plotted together with the (110) magnetic peak intensity at 0.5 K ($\blacksquare$). Vertical dashed lines delineate the range of fields where there are similar width steps in the Yb moment measured both by neutrons and by dc magnetization measurements. The magnetic field in both experiments is perpendicular to $c$ axis. \label{fig_neutrons}}
\end{figure}

The most direct information about the evolution of antiferromagnetic order with field and temperature comes from neutron diffraction measurements. We previously showed that the magnitude of the magnetic part of the (110) Bragg peak  in zero magnetic field obeys a mean field temperature dependence, consistent with the mean field character of the specific heat near $T_{\rm N}$~\cite{janssen2010}. Fig.~\ref{fig_neutrons}(a) confirms that magnetic field decreases the magnitude of the order parameter, and for temperatures larger than $\simeq 1.2$ K, it drops smoothly to zero along the antiferromagnetic phase line.  We have added these critical fields and temperatures to the phase diagram in Fig.~\ref{fig_HT_phase_diagram}, showing that they are in good agreement with values for $T_{\rm N}(B)$ obtained from specific heat and magnetization measurements. For $T \leq 1.2$ K, there is a distinct broadening of the transition, and at the lowest temperatures there is a pronounced step in the moment $\Delta M \simeq 0.2\mu_{\rm B}$/Yb centered at the critical field $B_{0}=1.85$ T. Like the step in $M(B)$, the breadth of the step in the ordered moment remains considerable, even at the lowest temperatures. Fig.~\ref{fig_neutrons}(b) shows that the transition widths found in the two experiments are very similar, $\simeq 0.25$ T.

\begin{figure}
\includegraphics[width=6.0cm]{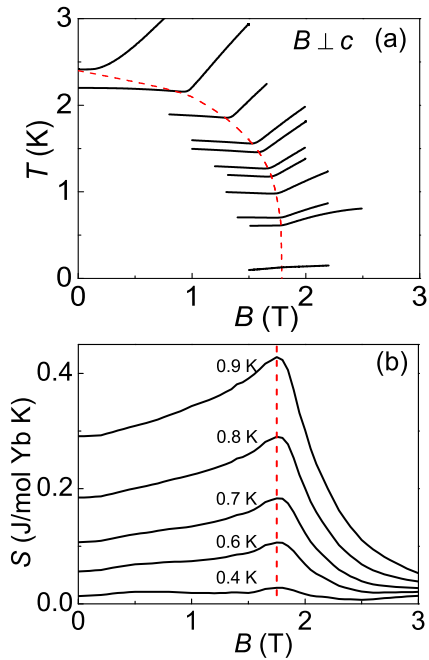}
\caption{(Color online) (a) The effect of magnetic fields perpendicular to the $c$ axis on different initial sample temperatures. The dashed line indicates the antiferromagnetic transition $T_{\rm N}(B)$ taken from Fig. 2, where the sample temperature increases due to the magnetocaloric effect. (b) Entropy $S$ calculated from specific heat $C_{\rm M}$ shown in Fig. 1 at different temperatures. Dashed line indicates that the maximum entropy occurs along the field independent antiferromagnetic phase line $T_{\rm N}(B)$.\label{fig_MCE}}
\end{figure}

The picture that emerges from the specific heat, magnetization, and neutron diffraction experiments is that the antiferromagnetic phase transition is continuous and  mean-field like in low fields, but when magnetic fields suppress  $T_{\rm N}$ to values less than $\simeq 1.2$ K, the broadened steps in the moment suggest that the transition may develop a first-order character. To test this hypothesis, we have carried out measurements of the magnetocaloric effect (MCE) to determine if a latent heat is associated with the antiferromagnetic transition along the vertical part of the phase line, i.e. when $T_{\rm N}\leq 1.2$ K. The MCE is the temperature change of a material when a magnetic field is changed adiabatically~\cite{tishin1999,pecharsky1999}, and it has been established in a number of correlated electron systems to be a practical and sensitive way to detect latent heat at a magnetic phase transition~\cite{bianchi2002,rost2009,jaime2002}. The results are shown in Fig.~\ref{fig_MCE}(a), where the solid line represents the sample temperature $T$, measured as the magnetic field is scanned.  A clear increase in the slope $dT/dB$ is observed as the antiferromagnetic phase is exited at $T_{\rm N}(B)$, but there is no discontinuity or jump in $T(B)$ anywhere along the phase line, either for $T_{\rm N}\geq 1.2$ K where the transition is definitively continuous, or at lower values where the nature of the transition is more ambiguous. We note that no differences are found along the phase line between increasing and decreasing field sweeps. Since the MCE measurements find that no latent heat is associated with the antiferromagnetic phase line in $\YP$, we conclude that the transition is continuous for all nonzero values of $T_{\rm N}$.

Since the MCE experiments approximate the adiabatic condition, the slope differences at $T_{\rm N}(B)$ found in Fig.~\ref{fig_MCE}(a) imply that the antiferromagnetic and paramagnetic states have different entropies, and that the difference between their respective entropies $\Delta S$ becomes increasingly small with reduced temperature.  This conclusion is supported by the field dependence of the entropy $S$, extracted from specific heat measurements (Fig.~\ref{fig_MCE}(b)), where we see a broad maximum in $S$ at $T_{\rm N}$ with a magnitude that decreases with decreasing temperature. Despite the steps observed in $M(B)$ and neutron diffraction experiments for $T\leq 1.2$ K, the MCE measurements apparently rule out a first order antiferromagnetic transition in $\YP$ for nonzero $T_{\rm N}$. Does this argument extend to $T_{\rm N}=0$ ?  The $Clausius-Clapeyron$ equation relates the slope of the antiferromagnetic phase line $dT_{\rm N}/dB$ to the differences between the magnetizations and entropies of the antiferromagnetic and paramagnetic phases at $T=0$: $dT_{\rm N}/dB=-\Delta M/\Delta S$.  The third law of thermodynamics requires that $\Delta S=0$ for $T=0$, and the vertical nature of the phase line $T_{\rm N}$ at the critical field $B_{\rm 0}$ implies that $dT_{\rm N}/dB \rightarrow -\infty$ for $T_{\rm N}=0$. The $Clausius-Clapeyron$ equation is satisfied at $T_{\rm N}=0$ when the transition is between two states with different magnetizations, i.e. $\Delta M \neq 0$, as we have seen in both the magnetization and neutron diffraction measurements. Our conclusion is that the antiferromagnetic phase line $T_{\rm N}(B)$ in $\YP$ is continuous at all nonzero temperatures, but terminates in a $T=0$ first order transition at a critical field $B_{\rm 0}=1.85$ T.

\section{Discussion and Conclusion}

\begin{figure}

\includegraphics[width=8.0cm]{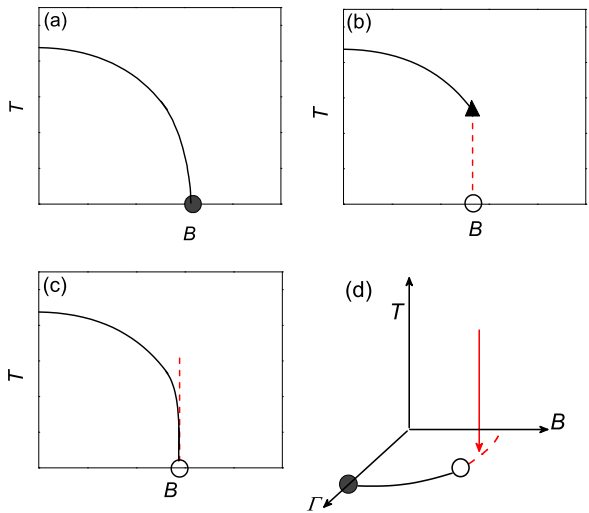}
\caption{(Color online) Schematic phase diagrams for field tuned antiferromagnets. (a) The phase line of a field tuned antiferromagnet remains second order at all fields (solid line), ending at a continuous transition with $T_{\rm N}=0$, i.e. a quantum critical point ($\bullet$). (b) The phase line of a field tuned antiferromagnet is initially second order(solid line), but this phase line terminates at a tricritical point($\blacktriangle$). For smaller values of $T_{\rm N}$, the phase line is first order (dashed line), ending at a first order transition where $T_{\rm N}=0$, i.e. a quantum critical endpoint ($\circ$). (c) An intermediate situation between (a) and (b), where the first order transition line in (b) has shrunk to a single point with $T_{\rm N}=0$, a quantum critical endpoint ($\circ$). For all nonzero values of $T_{\rm N}$, the phase line is continuous but strongly modified from the second order line shown in (a). (d) A three dimensional phase diagram with no magnetic order for $T>0$ at any value of field or other non-thermal variable $\Gamma$, such as pressure. There is a quantum critical endpoint ($\circ$) in the $T=0$ $B-\Gamma$ plane, that separates a first order line (dashed line) from a line of continuous transitions (solid line) that ends in a quantum critical point for $B=0$ ($\bullet$). The red vertical arrow indicates the effect of lowering temperature in a metamagnet, defined as a system that has no long-ranged order for $T\neq0$, but positioned in the $B-\Gamma$ parameter space close to a quantum critical endpoint($\circ$).  \label{fig_HT} }
\end{figure}

Our current understanding is that there is no universal path by which magnetic fields suppress antiferromagnetic order to zero temperature in heavy fermion compounds, and the schematic phase diagrams presented in Fig.~\ref{fig_HT} seek to categorize the simplest possibilities that have been identified by experiments. They are not meant to capture the full complexity of heavy fermion compounds, which may pass through a multiplicity of different structures en route to the collapse of magnetic order~\cite{stewart2001}, but rather to focus on the final phase line that separates magnetic order from the paramagnetic state.  To our knowledge, all heavy fermion antiferromagnets order via a continuous transition in zero field. Fig.~\ref{fig_HT}(a) depicts the situation found in systems like YbRh$_{2}$Si$_{2}$~\cite{custers2003}, YbPtIn~\cite{morosan2006}, CeCu$_{6-x}$Au$_{x}$~\cite{lohneysen2001}, and CeIn$_{3-x}$Sn$_{x}$~\cite{silhanek2006}, where the antiferromagnetic phase line remains continuous as $T_{\rm N}\rightarrow 0$ at a quantum critical field $B_{\rm QCP}$. Bulk properties such as the magnetization scale as functions of $T$ and ($B-B_{\rm QCP})$~\cite{gegenwart2005,custers2003}, and the magnetic Gr\"{u}neisen parameter diverges as well for $T=0$ and $B=B_{\rm QCP}$~\cite{zhu2003,kuchler2003}. Given that all experiments have a lower temperature limit, it is fair to say that it is not known in any compound whether the antiferromagnetic phase line is continuous to $T_{\rm N}=0$. However, it is evident that the scaling associated with the quantum critical point at $T_{\rm N}=0$ and $B=B_{\rm QCP}$ dominates many of the measured quantities over a wide range of fields and temperatures.

A very different situation is realized when magnetic fields are applied to conventional antiferromagnets such as  rare earth aluminum garnets and FeCl$_{2}$~\cite{stryjewski1977,birgenau1972,dillon1978}, which have continuous antiferromagnetic transitions in zero field (Fig.~\ref{fig_HT}(b)). Here, the antiferromagnetic phase line is initially second order, but terminates at a tricritical point~\cite{blume1974,giordano1976}.  Since magnetic order involves a broken symmetry, the phase line must continue to $T_{\rm N}=0$, and it does so as a line of first order transitions that terminate at a QCEP. Scaling is found in systems of this type, both of the conventional variety in low and zero fields, but more prominently in the vicinity of the tricritical point~\cite{dillon1978,shang1980}. This phase diagram is very similar to the one that was both predicted~\cite{belitz2005} and experimentally realized~\cite{pfleiderer2005} in field and pressure tuned metallic ferromagnets where disorder is weak. To our knowledge, the phase diagram in Fig.~\ref{fig_HT}(b) has been found only in ferromagnetic UGe$_{2}$~\cite{taufour2011}, and not in any antiferromagnetic heavy fermion compounds.

The phase diagram of Fig.~\ref{fig_HT}(c) represents a situation that is intermediate between Figs.~\ref{fig_HT}(a) and ~\ref{fig_HT}(b), in that the line of first order transitions has now shrunk to a single point at $T=0$, and it is the influence of this point that keeps the lowest temperature part of the phase line from becoming the more conventional second order phase line found in YbRh$_{2}$Si$_{2}$ (Fig.~\ref{fig_HT}(a)).  This is the phase diagram that best describes $\YP$, and perhaps as well Yb$_{5}$Pt$_{9}$~\cite{kim2006}, CeRh$_{2}$Si$_{2}$~\cite{knafo2010}, YbNiSi$_{3}$~\cite{Avila2004, Sergey2007} and CeNiGe$_{3}$~\cite{Mun2010}.  Here, the phase line is always continuous for $T_{\rm N}\neq 0$, and no latent heat is found anywhere along the phase line. The phase line superficially resembles the first order phase line of Fig.~\ref{fig_HT}(b), since it becomes vertical as $T_{\rm N}\rightarrow 0$. The initial stabilization of antiferromagnetic order as a second order transition at $B=0$ implies the general importance of long wavelength critical fluctuations through much of the $B-T$ plane, and the initial divergence of the susceptibility at the critical field where  $\chi(T)\sim T^{-x}$,  generally reflects these correlations. Since a true quantum critical point is ultimately avoided in systems described by the phase diagram in Fig.~\ref{fig_HT}(c), the longest wavelength fluctuations must either be absent, as in disordered systems, or are prohibited in some way from contributing to the physical observables.  We hypothesize that their absence is responsible for the breakdown of scaling near the QCEP, and for the general appearance of the phase line, which increasingly resembles a first order phase line, lacking only the latent heat. Ultimately the failure of universality as $T_{\rm N}\rightarrow 0$ causes the antiferromagnetic phase line to terminate in a first order phase transition at zero temperature $T_{\rm N}=0$, also known as a quantum critical end point.

The most unimpeded view of the properties of a quantum critical endpoint is found in systems in which no magnetic order is present, at least for $T\neq 0$. The most heavily studied examples of these so-called metamagnetic systems are CeRu$_{2}$Si$_{2}$ and Sr$_{3}$Ru$_{2}$O$_{7}$~\cite{weickert2010,flouquet2002,grigera2001}. The signature of metamagnetism is steps in the magnetization whose breadth decreases with decreasing temperature. In some cases, a full field-driven first order transition results below a certain onset temperature~\cite{rossat1985}, but for CeRu$_{2}$Si$_{2}$ and Sr$_{3}$Ru$_{2}$O$_{7}$ there is no sign of long-ranged magnetic order at any field or temperature. In both cases, there is a pronounced enhancement of the magnetization and specific heat near the critical field, and with reduced temperature the associated magnetic susceptibility begins to diverge as $\chi(T)\sim T^{-x}$~\cite{tautz1995}. Instead of a maximum in the specific heat, a dip is found in $C$ at the critical $B$. Unlike the case of $\YP$ where the termination of the nonzero temperature part of the phase line necessitates a true phase transition at $T=0$, no fine tuning is required for the metamagnets. All that is required is that the metamagnet is sufficiently close to a quantum critical endpoint,  accessible by tuning a nonthermal variable such as field angle in Sr$_{3}$Ru$_{2}$O$_{7}$~\cite{grigera2001}, or pressure in either system~\cite{flouquet1995,wu2011} (Fig.~\ref{fig_HT}(d)).

Unlike the case of clean ferromagnets, where it is theoretically and experimentally agreed that the phase line is initially continuous at small fields, but ultimately must become first order when the Curie temperature becomes sufficiently small, there is much less theoretical guidance for the range of behaviors that might be possible for antiferromagnets when $T_{\rm N}\rightarrow 0$. There is a continuing need to identify new systems that exemplify the differing phase diagrams that are represented in Fig.~\ref{fig_HT}. There are significant and intrinsic obstacles that make the search for such systems inherently challenging. One complication is that the suppression of magnetic order can enable the stabilization of competing collective phases, most notably superconductivity, as found in CeCoIn$_{5}$~\cite{paglione2003}, CeRhIn$_{5}$~\cite{park2008}, and CeCu$_{2}$Si$_{2}$~\cite{stockert2011}. However interesting and significant, these new phases obscure the part of the phase diagram where antiferromagnetic order vanishes. Similarly, experiments must be conducted at the very lowest temperatures to determine whether the quantum critical scaling is robust, or alternatively if universality fails and the antiferromagnetic transition becomes first order when $T_{\rm N}$ becomes sufficiently small.

Perhaps the most compelling aspect of the phase diagrams in Fig.~\ref{fig_HT} is their potential relationship to the underlying electronic structure. This has been studied extensively in the metamagnets, and in CeRu$_{2}$Si$_{2}$ magnetic fields are thought to drive a continuous evolution of the electronic structure from the $B=0$ limit where one of the spin polarized Fermi surfaces is favored in field, with the other vanishing at a Lifshitz transition at the metamagnetic field~\cite{daou2006}. In contrast to this case where the electrons are always delocalized, a rather different situation is realized in the heavy fermion YbRh$_{2}$Si$_{2}$~\cite{custers2003}.  Here the local moment character of the Yb moments is completely quenched near a Kondo temperature $T_{\rm K}$ that is  well in excess of the N\'{e}el temperature. Consequently, antiferromagnetic order must be considered to be a collective instability of the fully hybridized Kondo lattice, and magnetic fields drive a delocalization transition at the critical field $B_{\rm QCP}$ that is akin to a Mott transition, increasing the size of the Fermi surface~\cite{si2010}. Much of the $B-T$ phase diagram is affected by this transition, which coincides at $T=0$ with the antiferromagnetic quantum critical point in pure YbRh$_{2}$Si$_{2}$~\cite{gegenwart2008}, but remains a separate transition under Co and Ir doping~\cite{friedemann2009}. In contrast, the antiferromagnetic order that is found at the N\'{e}el temperature $T_{\rm N}$ in $\YP$ at zero field involves well localized Yb moments that are essentially unaffected by the Kondo effect, which we conclude occurs below a characteristic temperature $T_{\rm K}$ that is smaller than the ordering temperature itself, i.e. $T_{\rm K}\leq T _{\rm N}$~\cite{bennett2009,janssen2010}. The antiferromagnetic order is conventional, with a staggered Yb moment that is consistent with a doublet ground state~\cite{janssen2010}, and with spin waves that result from the exchange splitting of this state of the crystal electric field manifold~\cite{aronson2010}. It is tempting indeed to speculate that the very different natures of the Yb magnetism in YbRh$_{2}$Si$_{2}$ and $\YP$ may be responsible for their very different antiferromagnetic phase diagrams, represented in Figs ~\ref{fig_HT}a, and ~\ref{fig_HT}c, respectively.  Lacking a more comprehensive set of well characterized compounds with vanishing N\'{e}el temperatures, this association remains for now unproven.

To conclude, we have used measurements of the specific heat, magnetization, neutron diffraction, and magnetocaloric effect to establish the field-temperature phase diagram of the heavy fermion antiferromagnet Yb$_{3}$Pt$_{4}$. The antiferromagnetic transition is initially continuous in zero field, but magnetic fields applied in the easy ab plane reduce the N\'{e}el temperature in $\YP$  to zero temperature at a critical field $\simeq$ 1.85 T. The antiferromagnetic phase line becomes very steep  at low temperatures, and within the accuracy of our measurements becomes independent of field as $T_{\rm N}\rightarrow 0$. The appearance of the phase line is suggestive that the antiferromagnetic transition in $\YP$ becomes first order, however magnetocaloric effect measurements find no evidence for a latent heat for any value of $T_{\rm N}$. We conclude that the antiferromagnetic transition in $\YP$ is continuous, at least for $T_{\rm N}>0$. A step in the moment is observed at the critical field in both magnetization and magnetic neutron diffraction measurements, and the associated susceptibility $\chi=dM/dB$ at the critical $B$ initially increases with decreasing temperature, i.e. $\chi \sim T^{\rm -1/2}$, signifying that the step width is decreasing. However, the incipient divergence in $\chi$ is cutoff below $\simeq 0.35$ K, a behavior familiar from metamagnetic systems like CeRu$_{2}$Si$_{2}$ and Sr$_{3}$Ru$_{2}$O$_{7}$. Accordingly, we propose that the low temperature properties of $\YP$ are controlled by the quantum critical endpoint that is created when the antiferromagnetic phase line terminates at zero temperature.  These measurements position $\YP$ as one of the few  antiferromagnets from the heavy fermion class that do not seem to have true quantum critical points, formed when a second order phase transition is suppressed to zero temperature by magnetic field tuning. The field temperature magnetic phase diagram of $\YP$ seems to form a link between those of most field-tuned heavy fermions, which are dominated by a quantum critical point, and those of conventional magnetic insulators, where the central features are a tricritical point and a line of first order transitions terminating in a quantum critical end point.

\begin{acknowledgments}
We would like to thank Julia Scherschligt for much help with the $^3$He probe during the neutron scattering experiments in NIST NCNR. We would also like to thank Neil Dilley (Quantum Designs) and Andrea Candini for useful discussions regarding the Hall sensor magnetometer, and for providing prototypes. Work at Stony Brook University is supported by the National Science Foundation grant 0907457. The identification of any commercial product or trade name does not imply endorsement or recommendation by the National Institute of Standards and Technology.
\end{acknowledgments}


\end{document}